\documentclass[graybox]{svmult}

\usepackage{helvet}         
\usepackage{courier}        
\usepackage{type1cm}        
\usepackage{amsmath}    
\usepackage{amssymb}    
\usepackage{amsfonts}   

\usepackage{mathrsfs}   
\usepackage{bbm}        

\usepackage{makeidx}         
\usepackage{graphicx}        
\usepackage{multicol}        
\usepackage[bottom]{footmisc}

\usepackage{comment} 
\usepackage{color}

\def\tred{\textcolor{red}}
 
\def\cZ{\mathcal{Z}}

\def\bea{\begin{eqnarray}}
\def\eea{\end{eqnarray}}
\def\cO{ \mathcal{O}}
\def\nnm{\nonumber} 
\def\mC{ \mathbb{C}} 
\def\tr{ {\rm tr}}
\def\cA{ \mathcal{A}  } 
\def\cM{\mathcal{M}}
\def\sym{{\rm sym}}
\def\cN{ \mathcal{N}} 
\def\bV{ \overline{V} } 
\def\cK{ \mathcal{K}}

\begin{document}

\title*{ \hfill {\normalsize \normalfont  QMUL-PH-26-04}\\[1em]  Finite-dimensional algebras, gauge-string duality and thermodynamics. }
\author{Sanjaye Ramgoolam}
\institute{Sanjaye Ramgoolam \at Queen Mary University of London, Mile End Road, London, UK \\ \email{s.ramgoolam@qmul.ac.uk}}
%
%
\maketitle

\abstract{
Gauge-invariant polynomial functions of matrix and tensor variables capture  combinatorial structures of gauge-string duality, which can be usefully  organised using finite-dimensional associative algebras. I review recent work on  eigenvalue systems using these  algebras as state spaces, which provide efficient computational algorithms for the construction of  orthogonal bases for multi-matrix invariants. Algebraic counting formulae in matrix and tensor systems with $U(N)$ as well as $S_N$ symmetry have led to gauged quantum mechanical models which display a negative branch of specific heat capacity in the micro-canonical ensemble followed by positive specific heat capacity at larger energies measured by a polynomial degree parameter $n$. The negative branch is associated with near-exponential or factorial growth of degeneracies for $ n \gg 1$ in a region of large $N$ stability, while the positive branch occurs  when the finite $N$ reduction of degrees of freedom takes over as $n$ becomes sufficiently large compared to $N$. 
}

\section{Introduction}
\label{sec:intro} 

Gauge–string duality provides a remarkable framework in which gravitational physics in Anti–de-Sitter space can be studied using conformal field theory. In the planar limit, much is understood about correlators at both strong and weak coupling, and powerful integrability techniques are available. By contrast, at finite $N$  the Hilbert space acquires subtle $N$-dependent features, reflecting new constraints on gauge-invariant operators that are invisible in the planar limit.

I review two recent developments relevant to finite-$N$ physics in AdS/CFT. One concerns Casimir eigenvalue systems associated with the representation-theoretic organisation of multimatrix gauge-invariant operators \cite{PRS}. The other concerns thermodynamic behaviour arising from finite-$N$ effects, including negative specific heat in the microcanonical ensemble \cite{PIMQM-Thermo,PITQM}. Both are informed by families of finite-dimensional algebras associated with spaces of operators at fixed classical dimension.

In the archetypal example of the AdS/CFT correspondence \cite{malda,gkp,witten}, $\mathcal{N}=4$ super-Yang--Mills theory with gauge group $U(N)$ is conjectured to be dual to type IIB string theory on $\mathrm{AdS}_{5}\times S^{5}$, and an important strand of this correspondence concerns the emergence of ten-dimensional space-time from gauge-invariant variables built from $N\times N$ matrices. The finite-dimensional associative algebras that organise the combinatorics of such invariants provide an effective language for analysing this emergence, and serve as a bridge between representation-theoretic structures and thermodynamic behaviour indicative of gravitational physics.

Historically, combinatorial mechanisms behind gauge/string dualities have taken several forms: localisation to eigenvalues in matrix models (see e.g. the reviews \cite{Klebanov,GinsMoore}, the role of symmetric groups in two-dimensional Yang--Mills theory and its string expansion (e.g. reviews \cite{CMR}), and the representation theory of superconformal algebras and integrable spin chains in the spectrum of $\mathcal{N}=4$ SYM  (e.g. review \cite{Dobrev,Integrability}). From the point of view of local operators in the four-dimensional gauge theory, the simplest laboratory is provided by holomorphic polynomial invariants constructed from a single complex matrix $Z=X_{1}+iX_{2}$. The gauge-invariant quantities obtained by organising index contractions in terms of permutations are naturally classified by conjugacy classes of the symmetric group $S_{n}$, when the degree of the polynomial is $n$.  The sums of permutations in fixed conjugacy classes span the centre of the group algebra  $\mathbb{C}(S_{n})$.  Generalisations of these algebras are relevant to  polynomial functions of multiple matrix variables.

A key point is that these algebras do not simply parametrise invariants in a convenient way: they capture the finite-$N$ structure of correlation functions and provide a practical route to orthogonal bases of operators. The use of projectors labelled by Young diagrams with bounded height  $ l(R ) \le N $reflects the Schur--Weyl dualities arising from the action of $U(N)$ and the symmetric group $S_n$  on tensor powers $V_{N}^{\otimes n}$ 
\bea 
V_N^{ \otimes n } = \bigoplus_{ \substack { R \vdash n \\ l(R) \le N }  } V_R^{ U(N)} \otimes V_R^{ S_n } \nnm 
\eea
This description correctly tracks the counting of states both in the stable regime $ n \le N  $ 
as well as physically interesting regimes  $n \sim N$ relevant to giant gravitons \cite{mst,myers,hash,BBNS,CJR}  and $ n \sim N^2$ relevant to 
deformations of the $AdS_5 \times S^5$ geometry \cite{LLM}. This finite-$N$ control becomes crucial when we later examine the thermodynamic behaviour of models with many invariant degrees of freedom, particularly in regimes where the microcanonical temperature decreases with increasing energy---a signal of negative specific heat capacity  familiar from gravitational physics \cite{Lynden-Bell,HawkPage}  and arising naturally from matrix/tensor quantum thermodynamics \cite{PIMQM-Thermo,PITQM}

This contribution is organised as follows. Section 2 reviews the  parameterisation of one-matrix and multi-matrix gauge-invariants in $U(N)$ gauge theories with the help of permutations, building up to the use of permutation algebras in the construction of orthgonal bases for the free field inner product on the gauge invariants. Section 3 shows how structural properties of the permutation algebras can be used to construct algorithms for these bases. 
Section 4 explains how the algebraic perspective on the counting of gauge invariants in the context of $U(N)$ gauged quantum mechanical models was used to show that  a sector of  the quantum mechanics displays negative heat capacity. Section 5 shows that the same phenomenon occurs in models where matrix variables are replaced with higher tensor variables and where the continuous gauge symmetry $U(N)$ is replaced by the finite symmetric group gauge symmetry $S_N$.

\section{Permutation algebras for matrix invariants}
\label{sec:algebras}

The index structure of local operators in gauge theories with matrix-valued fields can be systematically organised in terms of permutations. We briefly recall this construction for one and two complex matrices, and emphasise the role of permutation conjugation in defining the associated algebras. Fourier transformation on these algebras relates combinatorial bases, defined directly using the conjugation operations, to representation theoretic bases labelled by Young diagrams. It  is a powerful source of physical information about the combinatorics of multi-matrix invariants for general values of the classical operator dimension at finite $N$.   This approach to matrix invariants has been instrumental in the study of giant gravitons and open strings attached to them. 

\subsection{Index contractions and permutation equivalence}

For a single complex matrix $Z$ of size $N$, holomorphic gauge-invariant polynomials of degree $n$ can be constructed  by using permutations $ \sigma \in S_n$ to parametrise index contractions 
\bea\label{FundLink}  
&& \cO_{ \sigma } ( Z )  =  \sum_{ i_1 , \cdots , i_n =1 }^{ N } Z^{ i_1}_{ i_{ \sigma ( 1) } }  Z^{ i_2 }_{ i_{ \sigma ( 2) } }  \cdots Z^{ i_n }_{ i_{ \sigma ( n) } } , ~~~~ \sigma \in S_n  \nnm 
\eea 
with equivalence under conjugation
\bea\label{equiv} 
O_{\gamma\sigma\gamma^{-1}}(Z) = O_{\sigma}(Z),
\qquad
\gamma \in S_n .
\eea
The polynomial invariants are therefore naturally parametrised by the conjugacy classes of $S_n$, and the vector space of invariants is isomorphic to the centre $\cZ(\mathbb{C}(S_n))$ for $n \le N$. Finite $N$ effects, in the regime $ n > N $ are captured by sub-algebras $ [ \cZ(\mathbb{C}(S_n))]_N$ defined using the  representation-theoretic projector basis of  $\cZ(\mathbb{C}(S_n))$.  This representation theoretic mechanism extends to multi-matrix and tensor systems, as we will describe shortly.

The map between permutations in $S_n$ and gauge-invariant operators can be extended by linearity  to the group algebra  $ \mC ( S_{n} ) $,  
\bea\label{BasicConn}  
&& \cO : \sigma \rightarrow \cO_{ \sigma } ( Z ) \cr 
&& \cO : \sum_{ \sigma } c_{ \sigma } \sigma \rightarrow  \sum_{ \sigma }  c_{ \sigma } \cO_{ \sigma } ( Z )  \nnm 
\eea
thus defining 
\bea\label{mapCSPol} 
\cO : \mC ( S_n) \rightarrow \hbox{ gauge invariant polynomials of degree $n$ in $Z$ } ~~~
\eea

The equality of invariant  polynomials  in \eqref{equiv}   corresponds to an equivalence relation on permutations 
\bea 
\sigma    \sim \gamma \sigma \gamma^{ -1} \nnm 
\eea 
While the permutations $ \sigma \in S_n  $ parametrises the index contractions, the permutations $ \gamma \in  S_n $ generate equivalences. In this case, the equivalence classes are conjugacy classes in $S_n$. 

The conjugacy classes are naturally associated to a sub-algebra of $ \mC(S_n)$. 
For each conjugacy class $C$, consider  the class sums $T_C$ defined as 
\bea 
T_{ C  } = \sum_{ \sigma \in C } \sigma \in \mC ( S_n )  \nnm 
\eea  
These commute with any permutation,  hence they belong to the centre  
\bea 
T_C \in ~~  \hbox { Centre of } ~~   \mC ( S_n )  \equiv  \cZ ( \mC ( S_n )   )  \nnm 
\eea 
It is also true that the centre is spanned by these class sums 
\bea 
\cZ ( \mC ( S_n )   )  = \hbox{ Span  }  \{ T_C : C \in \mathrm{ Set ~of ~conjugacy ~classes~ of}   S_n \}    \nnm 
\eea

The class sums obey an algebra of the form  
\bea 
T_C T_D =  \sum_{ E } \mathcal{C}_{ CD }^{  E } T_E \nnm 
\eea
with non-negative integer structure constants $ \mathcal{C}_{ CD }^{  E }  $. 
These appear in the correlation functions in matrix models. For example in the Gaussian complex one-matrix model, 
\bea 
{ 1 \over \cZ } \int  [ dZ  ] e^{ -  {  1 \over 2 }  \tr  ( ZZ^{ \dagger} )  } \cO_{ \sigma } ( Z ) \cO_{ \tau  } ( Z^{\dagger}  ) }   =  n! \sum_{ E }  \mathcal{C}_{ C  , D }^E   N^{ \hbox{ number of cycles in }   E   \nnm 
\eea 
This equation plays an important  role in the calculation of correlators in the half-BPS sector of $\cN=4$ SYM, e.g. in \cite{CJR}. A more refined use of the structure constants arises when we modify the Gaussian term to include a background (un-integrated) matrix $B$ : 
\bea
e^{ -  {  1 \over 2 }  \tr  ( ZZ^{ \dagger} )  }   \rightarrow e^{ -  {  1 \over 2 }  \tr  ( ZBZ^{ \dagger} )  }  \nnm
\eea 
Correlators now take the form 
\bea 
\langle \cO_{ C } ( Z ) \cO_{ D } ( Z^{ \dagger} ) \rangle_B = n !
 \sum_{ E } \mathcal{C}_{ C  , D }^E   \cO_{ E } ( B ) \nnm 
\eea 
The original  derivation of  this equation and its multi-matrix/tensor generalisations is in \cite{RS23}. 

In addition to the centres of symmetric group algebras $ \cZ(\mathbb{C}(S_n))$ , the connection between permutations and invariants in \eqref{FundLink} motivates interest in  sub-algebras  $\big[ ~ \cZ (\mC( S_n)  ) ~ \big]_{ N }$ which depend on $N$, the  size of the matrix $Z$. We are thus led to infinite families of finite algebras 
\bea 
\bigoplus_{ n =0}^{ \infty }   \big[ ~ \cZ (\mC( S_n)  ) ~ \big]_{ N }  \nnm 
\eea
which are important for 1-matrix correlators. The parameters have the following significance : 
\bea 
  n &=&  \hbox{ degree of the polynomial in} ~~ Z \cr 
&=&  \hbox{ scaling dimension of the local operator in } ~~CFT4 \cr
& =&  \hbox{ energy of states in } ~~ AdS_5 \nnm 
\eea
There is interesting physics of giant gravitons associated with $ n \sim N , n \gg N $ as proposed  in \cite{CJR}.

\subsection{Representation-theoretic bases }

The algebra $\cZ(\mathbb{C}(S_n))$  admits a  natural basis labelled by Young diagrams. For $\cZ(\mathbb{C}(S_n))$, the standard projectors
\[
P_R = \frac{d_R}{n!}\sum_{\sigma\in S_n} \chi_R(\sigma)\,\sigma
\]
obey $P_R P_S = \delta_{RS} P_R$, with $R,S$ are Young diagrams with $n$ boxes. Their images under the map $\cO $ in \eqref{mapCSPol} form orthogonal bases of holomorphic operators at finite $N$, when the height condition $\ell(R)\le N$ is imposed. This is a consequence of the Schur-Weyl  decomposition of $V_N^{\otimes n}$ : 
\bea 
V_N^{ \otimes  n } = \bigoplus_{ \substack {  R \vdash n \\  l(R ) \le N  } }  V_R^{(U(N)} \otimes V_R^{ (S_n)} \nnm 
\eea
The finite sub-algebra $ \big[\cZ (\mC( S_n)\big]_{ N }$  is defined as  
the linear span (over $\mC$) of the projectors $P_R$ with $ l(R) \le N$. 
Young diagrams which have order one long columns (of length $l(R) \le N $) correspond to sphere giant gravitons, while those which have order one long rows of 
length comparable to or greater that $N$ correspond to dual AdS giants \cite{CJR}.

\subsection{ 2-matrix case : contraction permutations, equivalences and algebras  } 

For two matrices $Z$ and $Y$, and polynomial  degrees $(m,n)$, the invariants take the  form $O_{\sigma}(Z,Y)$ with $\sigma\in S_{m+n}$ : 
\bea 
\cO_{ \sigma }  ( Z , Y ) = \sum_{ i_1 , \cdots , i_{ m+n}  =1 }^{ N }   Z^{ i_1}_{ i_{ \sigma(1) } } \cdots Z^{ i_{m} }_{ i_{ \sigma (m) }} Y^{ i_{m+1} }_{ i_{ \sigma (m+1) }}  Y^{ i_{ m+n} }_{ i_{ \sigma (m+n)} } \nnm 
\eea
Thus there is a map from $ \mC ( S_{ m+n } ) $ to two-matrix invariants 
\bea 
\cO^{ (m,n)} : \sigma \rightarrow \hbox{ two-matrix invariants  of degree $(m,n)$ } 
\eea
Distinct permutations $ \sigma $ related by conjugation with $ \gamma \in S_m \times S_n \subset S_{m+n} $ map to the same gauge invariant polynomial. This leads to equivalence classes in $S_{m+n}$ generated by the subgroup conjugations. Each equivalence class defines a basis  element for a subalgebra  $\cA(m,n) $ of 
$ \mC ( S_{ m+n}] ) $ defined by 
\[
\cA(m,n)
 = \big\{ x\in\mathbb{C}(S_{m+n}) \; \big|\;
         gx = xg \ \text{for all } g\in S_m\times S_n \big\}.
         \] 
$\cA ( m , n ) $ is a non-commutative associative  algebra. It is also semi-simple, i.e. has a non-degenerate pairing inherited from $\mC ( S_{ m+n} ) $. In the parent algebra $\mC ( S_{ m+n} ) $ this is essentially the inner product where permutations form an orthonormal basis. As a result the general structures from the mathematical studies of semi-simple non-commutative associative algebras, notably the Wedderburn-Artin matrix decomposition,  are applicable, as detailed for example in \cite{CurtRein}. Additionally physical constructions such as symmetric group Clebsch-Gordan coefficients and branching coefficients as described in mathematical physics books, e.g. \cite{Hamermesh}, provide complementary insights.

\subsection{ Rep theory bases : 2-matrix case }

Analogous to the projector basis of $ \cZ ( \mC ( S_n) ) $ there is a 
basis of $\cA(m,n)$ 
\[
Q^{R, R_1,R_2}_{ \mu,\nu}
\]
labelled by triples $(R,R_1,R_2)$ with
\bea 
&& R\vdash (m+n) :  \hbox{ $R$ is a partition of $(m+n) $ } \cr 
&& R_1\vdash m :  \hbox{ $R_1$ is a partition of $m$ } \cr 
&& R_2\vdash m :  \hbox{ $R_2$ is a partition of $n$ } \cr 
&& 1 \le  \mu,\nu \le g(R_1,R_2;R) = \hbox{ Littlewood-Richardson coefficient for the triple}  \nnm 
\eea  
Thus 
\bea 
&&  \cA( m,n)  = \hbox{ Span } \{ Q^{R,R_1,R_2}_{ \mu,\nu} : R \vdash (m+n ) , R_1 \vdash m , R_2 \vdash n , 1 \le \mu , \nu  \le g ( R_1 , R_2 , R )    \} \cr 
&& 
\eea 

These are called {\it matrix units}  in the mathematical literature and can be constructed using group-theoretic matrix elements of permutations in the irreducible representation $R$ of $S_{m+n}$ along with 
 branching coefficients for reduction $ R \rightarrow R_1 \otimes R_2 $ from $S_{m+n}$ to the sub-group $S_m \times S_n$. This has been studied in the physics literature and used to construct bases for multi-matrix gauge invariants, which are orthogonal under the inner product defined by free field two-point functions \cite{KR,BHR1,BHR2,BCD,BDS,EHS}.  The  construction of multi-matrix bases drew on ideas developed in the context of  open strings attached to giant gravitons \cite{BHN,BBFH}.

A basis for the  finite $N$ space of matrix invariants is obtained by 
restricting  $\ell(R)\le N$. The algebra $\cA( m,n) $ was somewhat implicit these papers which focused on multi-matrix gauge invariants, and was made explicit as examples of permutation centraliser algebras in \cite{PCA}. 
The finite $N$ truncation of $ \cA ( m,n ) $ is a sub-algebra $  [ \cA( m,n) ]_N $ 
\bea\label{finiteNsubalg}  
[ \cA( m,n) ]_N = \hbox{ Span } \{ Q^{R,R_1,R_2}_{ \mu,\nu} : \ell ( R) \le N \} 
\eea
The map $ \cO^{ (m,n)} $ gives an isomorphism from $[  \cA ( m, n) ]_N $ to the linear space of polynomial gauge invariants of two matrices of size $N$, of degree $(m,n)$ in the two matrices,  where both matrices transform in the adjoint of the gauge group $U(N)$. 

The algebras $\cA (  m,n )  ,  [ \cA ( m,n) ]_N$ are central to recently developed efficient  computational  algorithms for finite-$N$ operator bases in multi-matrix problems \cite{PRS}. These bypass the explicit calculation of branching coefficients and Clebsch-Gordan coefficients for symmetric groups and are instead based on eigenvalue systems defined on the algebras $ \cA ( m ,n ) $, and which are  implemented using group-theoretic software such as GAP and sagemath  \cite{GAP}\cite{sagemath}.   We outline the algorithms in somewhat more detail in the next section. 

Developments based on the matrix units of $ \cA ( m,n)$  and related to open strings attached to giant gravitons are reviewed in \cite{RobRev,BerRev}.

\section{ Algorithms for orthogonal bases } 
\label{sec:algorithms}

In this section, we explain how the algebra $ \cA(m,n)$ described in section \ref{sec:algebras} leads to the efficient construction of orthogonal bases in multi-matrix systems.

Given any $ \sigma  \in \mC ( S_{ m+n} ) $ we can form the sum 
\bea 
\bar \sigma = \sum_{ \gamma \in S_m \times S_n }   \gamma \sigma \gamma^{-1} \nnm 
\eea
These sums commute with $ S_m \times S_n$ 
\bea 
\mu \bar \sigma = \bar \sigma  \mu ~~\hbox{ for } ~~\mu \in S_m \times S_n \subset S_{ m+n} \nnm 
\eea
and thus live in $ \cA ( m,n)$. 
A combinatorial basis for $ \cA ( m, n ) $  is given by  such sums  $ \bar \sigma_a$,  where $a$  labels orbits in  the group $S_{ m+n} $ generated by conjugation with elements of the sub-group $S_m \times S_n$.

\subsection{ The eigenvalue equations in $ \cA ( m ,n ) $ } 

Given partitions  $p,q,r$ of $(m+n), m , n$ respectively, which correspond to   cycle-structures in  $S_{m+n} , S_m , S_n$, there are central cycle sums 
\bea 
 T^{(m+n)}_{ p }  \in \cZ ( \mC ( S_{ m+n} ) )   \cr 
  T^{(m)}_{ q }  \in \cZ ( \mC ( S_{ m} ) ) \cr 
    T^{(n)}_{ r } \in \cZ ( \mC ( S_{ n} ) ) \nnm 
\eea 
The matrix units of $ \cA ( m , n ) $ satisfy  
\bea 
&& T_{ p}^{ (m+n)} Q^{R}_{ R_1 , R_2 , \mu , \nu }  =  { \chi^R ( T_p^{ (m+n)}  ) \over d_R     } Q^{R}_{ R_1 , R_2 , \mu , \nu }  \cr 
&& T_{ p}^{ (m)} Q^{R}_{ R_1 , R_2 , \mu , \nu }  =  { \chi^{R_1}  ( T_q^{ (m)}  ) \over d_{R_1}      } Q^{R}_{ R_1 , R_2 , \mu , \nu }  \cr 
&& T_{ r}^{ (n)} Q^{R}_{ R_1 , R_2 , \mu , \nu }  =  { \chi^{R_2}  ( T_r^{ (n)}  ) \over d_{R_2}  } Q^{R}_{ R_1 , R_2 , \mu , \nu } \nnm 
\eea
The eigenvalues are normalised characters which can be calculated from symmetric group character tables, and are in fact integers.

Matrices for the central elements in the combinatorial basis can be constructed 
\bea 
T_{ P }^{ (*) } \bar \sigma_a =   \sum_{ b } ( \cM_{ P } )_{ ba  } \bar \sigma_b \nnm 
\eea
where $P$ is one of the partitions $ p,q,r $ above. The matrix elements 
$ ( \cM_{ P } )_{ ba  } $ are also integers because of their combinatorial definition. We are thus led to look for null vectors of  integer matrices 
\bea 
 \left ( \cM_{ P }  - { \chi^{R_* } ( T_{P} ) \over d_{ R_* }  }  \right )_{ ba  }  \nnm 
\eea
There are  $g(R_1 , R_2 ; R )^2$ null vectors in $ \cA ( m ,n ) $  for every triple $ ( R , R_1 , R_2 )$. This has been coded using  software for group theory and algebra \cite{GAP,sagemath}. The codes are available in the accompanying files of the arXiv submission of \cite{PRS}.

\subsection{ Efficiency of algorithms and  minimal generating sets for $ \cZ ( \mC ( S_n) ) $. }

To distinguish the Young diagrams of $ S_n$, it suffices to choose a few cycle conjugacy classes
$ T_2 , T_3 , \cdots T_{ k_* (n) } $, where the permutations have a single non-trivial cycle of length $2,3, ..$.  $k_*(n) $. The computational study of character tables shows that $k_*(n)$  is an  integer that grows slowly with $n$ \cite{KempRam}.  For $ n $ up to $14$ it suffices to   use $ T_2  , T_3 $. 
While $T_{ P } $ for general $P \vdash n $, form a linear basis for $ \cZ ( \mC ( S_n ))$, these cycle classes form a non-linear generating set. Any projector in $ \cZ ( \mC ( S_n ))$ can be expanded as 
\bea 
P_R = \sum_{ \vec n } c^R_{ \vec n } \prod_{ a \in \{ 2 , \cdots , k_*(n) \} } T_a^{ n_a  } \nnm 
\eea
for appropriate coefficients  $c^R_{ \vec n }$ \cite{KempRam}. The relation between non-linear  minimal generating sets of conjugacy  classes and minimal sets of normalised irreducible characters which distinguish irreducible representations is extended to general groups in \cite{RS}. It has been used in complexity theoretic studies of centres of group algebras and character tables in 
\cite{QMRibb,ProjDetect,RowCol,PRSe}. 

An important aspect of these algorithms is that 
$N$ can be treated as a parameter. The complexity is controlled by $ ( m , n ) $, so we have manageable algorithms for $m+n$ up to around $15$, while $N$ can be arbitrary. This thus goes well beyond standard large $N$ expansion techniques.

\section{  Representation bases, counting and thermodynamics   }
\label{sec:thermodynamics}

In this section, we describe how counting formulae related to the algebraic finite $N$ bases for 
multi-matrix and tensor systems with $U(N)$ gauge symmetry lead to a  negative specific heat capacity branch of the energy-temperature curve in the micro-canonical ensemble. This is  of interest in the context of small black hole physics in AdS/CFT \cite{Hanada,Berenstein2018}. The transition from the negative to the positive branch occurs near a cross-over in the finite $N$ canonical ensemble, where the
specific heat capacity remains positive as expected on general grounds. 
 These observations extend to gauged quantum mechanical systems with $S_N$ symmetry, where the phenomenon occurs already for the 1-matrix problem - this is related to the very rapid growth of invariants as a function of polynomial degree when the gauge symmetry is reduced from $U(N) $ to $S_N$.

\subsection{ Micro-canonical negative specific heat capacity in a sector of the 2-matrix system}

The dimension of the state space of finite $N$ matrix invariants can be read off from  from representation basis \eqref{finiteNsubalg} 
\bea\label{dimRep} 
\hbox{ Dimension of }  [ \cA (  m, n) ]_{ N }  = 
\sum_{ \substack {  R \vdash (m+n ) \\ \ell (R ) \le N } } \sum_{ R_1 \vdash m } \sum_{ R_2 \vdash n } 
 ( g ( R_1 , R_2 , R ) )^2  \nnm 
\eea
The range  $ (m+n) \le N $ is called the stable range counting. For this range, as $N$ varies subject to the constraint, the dimension is unchanged.

The 2-matrix counting and basis problem considered above  has two distinct realisations in $ \cN=4$ SYM. When the two matrices are two complex matrices and we are considering holomorphic functions of the two matrices, this is the space of quarter-BPS operators at zero coupling, e.g. as considered in \cite{BHR1} (see also refs. therein). When there is just one complex matrix, and we consider polynomial functions which are not necessarily holomorphic, i,e we have multi-traces involving $ Z , Z^{ \dagger} $. The case $ m = n $ with equal numbers of  $ Z , Z^{\dagger} $ is  a sector that can be interpreted in terms of giant  graviton brane/anti-brane systems \cite{KR} as a zero charge sector. The counting for this zero-charge sector is thus 
\vskip.2cm 
\bea 
\cZ ( n ; N  ) =   \sum_{  \substack{  R \vdash (2 n )  \\ l(R) \le N } } \sum_{ R_1 \vdash  n } \sum_{ R_2 \vdash n } 
 ( g ( R_1 , R_2 , R ) )^2  \nnm 
\eea

This counting function gives the entropy in the micro-canonical ensemble 
\bea 
\hbox{ Entropy }  = S ( n  ; N ) = \log ( \cZ ( n ; N  ) )  \nnm 
\eea
Using the standard thermodynamic relation 
\bea 
dU =   T dS  \nnm 
\eea 
the micro-canonical temperature can be defined, in this case of discrete energy levels, as 
\bea 
T^{-1}_{ \rm micro }   = {  \partial S \over \partial n }  = S ( n+1  ; N )  - S ( n ; N ) \nnm 
\eea 
The energy-temperature curve shows negative specific heat capacity  branch at low energies,
 which turns turns around to give positive specific heat capacity for $n$ sufficiently large compared to $N$: see the figure \ref{EvsTmicroNeq13ZerCharge2Mat} from 
  \cite{PIMQM-Thermo}. 
The specific heat capacity remains positive in the canonical ensemble, and there is a Hagedorn transition similar to the one  in the full 2-matrix problem, without the zero charge $m=n$ constraint \cite{Minwalla,Sundborg}.  An important difference is that the constrained system has negative specific heat capacity, which is due the mildly sub-exponential  growth 
\bea 
\cZ ( n ) \sim {  4^n \over \sqrt { n } } 
\eea 
in the large $n$ stable regime of the counting \cite{RWZ}.  
The full 2-matrix counting is purely exponential in this asymptotic regime and has no negative specific heat capacity. 
\begin{figure}
\includegraphics[scale=0.5]{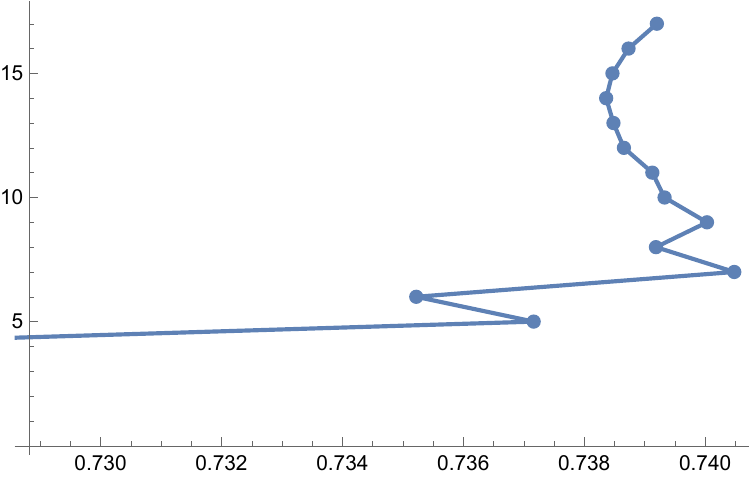} 
\caption{Micro-canonical energy versus temperature for zero charge complex matrix system $ N = 13$ with $ k $ equals $3$ to $18$ using the symmetric $D_{ \sym}$ discrete derivative. }  
\label{EvsTmicroNeq13ZerCharge2Mat}  
\end{figure}

\subsection{ Finite algebras and thermodynamics  for tensor invariants }

Consider invariants constructed from  complex 3-index tensors 
 $ \Phi_{ i j k } , \bar \Phi^{ i j k } $ transforming as a tensor product of 
 fundamental representations of $ U(N) \times U(N) \times U(N)$ \cite{BGSR}  : 
 \bea 
&&  \Phi \in V_N \otimes V_N \otimes V_N \cr 
&& \cr 
 &&  \bar \Phi \in \overline{ V}_N \otimes \bV_N \otimes \bV_N \nnm 
 \eea
Polynomials in $ \Phi , \bar \Phi $ can only be gauge-invariant if the degree in $\Phi $ is equal to the degree in $ \bar \Phi$. Let $n$ be this degree.

For general $n$, the invariants can be constructed from a triple of permutations 
\bea 
( \sigma_1  , \sigma_2 , \sigma_3 ) \in S_n \times S_n \times S_n \nnm 
\eea
 Equivalences generated by $ \gamma_L , \gamma_R \in S_n$ arise  from bosonic statistics   of $ \Phi , \bar \Phi $ 
\bea 
( \sigma_1 , \sigma_2 , \sigma_3  ) \sim ( \gamma_L  \sigma_1 \gamma_R , \gamma_L  \sigma_2 \gamma_R , \gamma_L  \sigma_3 \gamma_R  ) \nnm 
\eea
This gives a powerful approach for the counting of tensor invariants \cite{BGSR,BGSR2}.

The equivalence classes of permutations span an algebra  $ \mathcal{ K } ( n ) $ 
\bea 
{ \rm Dimension ~~ of ~~ algebra } ~~  \mathcal { K }  ( n ) = \sum_{ p \vdash n } | Aut ( p ) |  \sim n! \hbox{ at large $n$ }  \nnm 
\eea
The factorial growth and associated all orders asymptotics was studied in \cite{AsympTens}. 
The entropy $ S ( n ) = \log ( n! ) $ leads directly to negative specific heat capacity in the micro-canonical ensemble.  This much faster growth means that Hagedorn temperature vanishes in the large $N$ limit since 
\bea 
\sum_{ n }^{ \infty }  e^{ - \beta n + n \log n - n } \nnm 
\eea 
diverges for all finite $ \beta = 1/T$. The radius of convergence in temperature space vanishes.  
This was identified as a zero-temperature Hagedorn transsition in \cite{Tseyt,Kleb}. 

Finite $N$ effects lead to a sub-algebra 
$ [ \cK (n) ]_N $, which can be described in terms of a representation theory basis of $ \cK(n)$. 
The 
dimension of space of invariants  for general $N$, in terms of Kronecker coefficients is  
\bea 
\cZ ( N , n ) =  \dim \left (  [ \cK (n)  ]_N  \right ) 
\sum_{ \substack{ R , S , T \vdash n \\ \ell( R ) , \ell(S) , \ell(T) \le N }  } ( C ( R , S , T ) )^2 \nnm 
\eea 
Using this group-theoretic formula, the micro-canonical Energy-temperature curve shows a negative specific heat capacity branch expected from the large $n$ asymptotics in the stable range ( $n \le N$). This is followed by a positive SHC branch as the energy, equivalently the polynomial degree $n$ becomes sufficiently large compared to $N$ (see figure \ref{EvsTmicroNeq4k3to12} from \cite{PIMQM-Thermo}).

\begin{figure}
\includegraphics[scale=0.5]{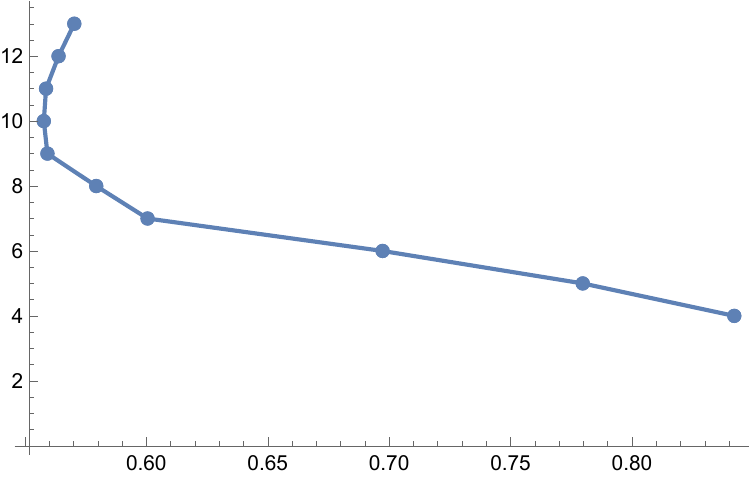}
\caption{Micro-canonical energy versus temperature for $3$-index tensor  $ N = 4$ with $ k $ equals $3$ to $12$ using the symmetric $D_{ \sym}$ discrete derivative.}  
\label{EvsTmicroNeq4k3to12}  
\end{figure}

\subsection{Partition functions and phase  structure in matrix and tensor models with $S_N$ gauge symmetry  }

Consider functions of  a real $ N \times N $ matrix with matrix elements 
$ X_{ ij} $, invariant under 
\bea 
X \rightarrow U X U^{ T }, \nnm 
\eea
where $U$ is a general permutation matrix in $S_N$. 

In the  canonical ensemble with $ x = e^{ - \beta } $, the partition function \cite{PIMQM-PF} is 
\bea 
\cZ ( N , x  ) = \sum_{ p \vdash N } \cZ (  N  , p ,  x ) \nnm 
\eea 
where $p$ is a partition of $N$ parametrised as 
$ p = [ a_1^{ p_1} , a_2^{ p_2} , \cdots , a_K^{ p_K } ] $ ; $ p_i\ge 1 ; \sum_{ i } a_i p_i =N$ and 
\bea\label{matpart}
\cZ (  N , p ,  x ) =  \prod_{ i, j \in \{ 1 , \cdots , K \} } 
{ 1 \over ( 1 - x^{ LCM (a_i , a_j ) } )^{ a_i a_j  p_i p_j \over LCM ( a_i , a_j ) } }   
\eea
This leads to similar thermodynamics in the micro-canonical ensemble, with a negative specific heat capacity branch followed by positive specific heat at higher energies as shown in Fig. \ref{MicroTempVsEnergy} from \cite{PIMQM-Thermo}.

 A similar phase structure occurs in the  $s$-tensor model with $S_N$ gauge symmetry \cite{PITQM}. The matrix model partition function formula \eqref{matpart}  generalises to 
\bea 
\cZ (  N  , p ,  x ) =  \prod_{ \{  i_1 , i_2 , \cdots , i_s \}  \in \{ 1 , \cdots , K \} } 
{ 1 \over ( 1 - x^{ LCM (a_{i_1}  , a_{i_2} , \cdots  ,a_{ i_s}   ) } )^{ a_{ i_1 } \cdots a_{ i_s} p_{ i_1} \cdots p_{ i_s}  \over LCM ( a_{i_1} \cdots , a_{ i_s} )     }    } \nnm 
\eea

As we have seen,  Schur-Weyl duality relates finite $N$ combinatorics for $U(N) $ gauge theories to symmetric groups $S_n$ for operators involving $n$ fields. Analogously for $S_N$ gauge theories the dual algebras for the case of degree $k$ operators are the partition algebra $P_k(N)$. This has been used to study large $N $ factorisation and orthogonal bases of operators in \cite{BPR1,BPR2}. The gauged quantum mechanical models have a path integral formulation which allows the physical derivation of the Molien-Weyl formula \cite{PIMQM-PI}. This formula is a powerful  tool  for obtaining the generating functions above \cite{PIMQM-PF,PITQM}. 

\begin{figure}
\includegraphics[scale=0.8]{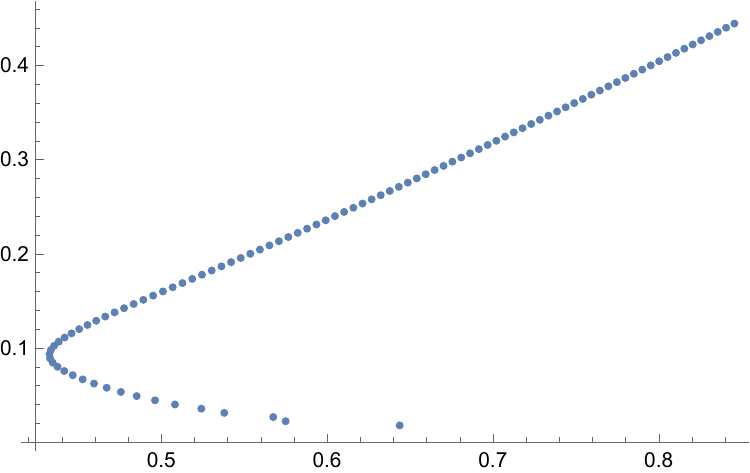} 
\caption{ Plot of micro-canonical energy $E={k \over N^2 }$ versus micro-canonical temperature at $N =15 $ for $k_{ min } = 4 , k_{ max} = 100 $ - using descending derivative -- produces consistent negative SHC trend below critical E  }
\label{MicroTempVsEnergy}  
\end{figure}

\section{ Summary }

 Polynomial functions of degree $n$  of matrix and tensor variables of size $N$, invariant under symmetries $U(N) $, arise in the classification of local operators in $ \cN=4$ SYM of fixed classical dimension $n$. Symmetric groups $S_n$ of permutations arise in parameterising the index contractions,  and associated sub-algebras of the group algebra $ \mC ( S_n)$ are very powerful tools in understanding the counting and correlators at finite $N$. We reviewed recent results on algorithms for the construction of operators and suprising thermodynamic features arising in the context of multi-matrix operators. Similar algebras and associated algorithms and thermodynamics arise when we go from multi-matrix systems to tensor systems, and when the gauge group is changed from $U(N)$ to $S_N$. 
 
The focus in this short review, in the case of $U(N)$ gauge group, has been to use permutations to organise the index contractions. Alternatively, by using transposes of the form $ Z^{ \otimes m } \otimes (Y^{T})^{\otimes n  } $  for two matrices or $ Z^{ \otimes m } \otimes (Z^*)^{\otimes n  } $ for a complex matrix and its conjugate, the contractions are organised by walled Brauer algebras $B_N(m,n)$ \cite{KR}  and orthogonal multi-matrix bases can also be organised by Brauer representation theory. This amounts to diagonalising a different set of charges in the free theory \cite{EHS}. The connection between $B_N(m,n)$ and tensor space operators has also found powerful uses in quantum information theory (e.g. \cite{WBAPBT}). The convergence of physical perspectives on $B_N(m,n)$ has led to new results on the non-semi-simple regime of the representation theory ($m+n > N$) where surprising links to two dimensional free field counting functions have been found \cite{SHOBNS}. 

Future directions include the application of this algebraic approach in the computation of loop corrections to classical dimensions and  the coupling of the matrix/tensor systems with negative specific heat capacities to external probes to investigate analogies to the physics of small black holes.

\vskip.2cm

\begin{acknowledgement}

I thank  my collaborators Denjoe O'Connor, Adrian Padellaro, Ryo Suzuki for the enjoyable collaborations \cite{PRS,PIMQM-Thermo,PITQM}  which led to the work presented here. I also thank  Joseph Ben Geloun,  Garreth Kemp, Yang Lei, Micha\l ~ Studzi\'nski for interesting related conversations. This work was supported by a Visiting Professorship at the Dublin Institute for Advanced Studies in 2024 and an ongoing Royal Society International Exchanges grant IEC-NSFC-242376 held jointly with Yang Lei at Soochow University, China.  I am grateful to the organizers of the ``XVI International Workshop LIE THEORY AND ITS APPLICATIONS IN PHYSICS'' (held in Varna in June 2025 and dedicated to the memory of Prof Ivan Todorov)  for the invitation to give a  presentation. 

\end{acknowledgement}


\begin{thebibliography}{99}



\bibitem{PRS}
A.~Padellaro, S.~Ramgoolam and R.~Suzuki,
``Eigenvalue systems for integer orthogonal bases of multi-matrix invariants at finite $N$,''
J.\ High Energ.\ Phys.\ \textbf{2025}, 111 (2025),
arXiv:2410.13631 [hep-th].

\bibitem{PIMQM-Thermo}
D.~O'Connor and S.~Ramgoolam,
``Permutation invariant matrix quantum thermodynamics and negative specific heat capacities in large $N$ systems,''
J.\ High Energ.\ Phys.\ \textbf{2024}, 161 (2024),
arXiv:2405.13150 [hep-th].

\bibitem{PITQM}
D.~O'Connor and S.~Ramgoolam,
``Gauged permutation invariant tensor quantum mechanics, least common multiples and the inclusion--exclusion principle,''
to appear, arXiv:2506.18813 [hep-th].


\bibitem{malda}
J.~M.~Maldacena,
``The Large $N$ limit of superconformal field theories and supergravity,''
Adv. Theor. Math. Phys. \textbf{2} (1998), 231–252,
arXiv:hep-th/9711200.

\bibitem{gkp}
S.~S.~Gubser, I.~R.~Klebanov and A.~M.~Polyakov,
``Gauge theory correlators from noncritical string theory,''
Phys. Lett. B \textbf{428} (1998), 105–114,
arXiv:hep-th/9802109.

\bibitem{witten}
E.~Witten,
``Anti de Sitter space and holography,''
Adv. Theor. Math. Phys. \textbf{2} (1998), 253–291,
arXiv:hep-th/9802150.


\bibitem{Klebanov}
I.~R.~Klebanov,
``String theory in two-dimensions,''
arXiv:hep-th/9108019.

\bibitem{GinsMoore}
P.~H.~Ginsparg and G.~W.~Moore,
``Lectures on 2-D gravity and 2-D string theory,''
arXiv:hep-th/9304011.

\bibitem{CMR}
S.~Cordes, G.~W.~Moore and S.~Ramgoolam,
``Lectures on 2-d Yang-Mills theory, equivariant cohomology and topological field theories,''
Nucl. Phys. B Proc. Suppl. \textbf{41} (1995), 184–244,
arXiv:hep-th/9411210.

\bibitem{Dobrev}
V.~K.~Dobrev,
``AdS / CFT correspondence and supersymmetry,''
arXiv:hep-th/0207116.

\bibitem{Integrability}
N.~Beisert \textit{et al.},
``Review of AdS/CFT Integrability: An Overview,''
Lett. Math. Phys. \textbf{99} (2012), 3–32,
arXiv:1012.3982.


\bibitem{mst}
J.~McGreevy, L.~Susskind and N.~Toumbas,
``Invasion of the giant gravitons from Anti-de Sitter space,''
JHEP \textbf{06} (2000), 008,
arXiv:hep-th/0003075.

\bibitem{myers}
M.~T.~Grisaru, R.~C.~Myers and O.~Tafjord,
``SUSY and Goliath,''
JHEP \textbf{08} (2000), 040,
arXiv:hep-th/0008015.

\bibitem{hash}
A.~Hashimoto, S.~Hirano and N.~Itzhaki,
``Large branes in AdS and their field theory dual,''
JHEP \textbf{08} (2000), 051,
arXiv:hep-th/0008016.


\bibitem{BBNS}
V.~Balasubramanian, M.~Berkooz, A.~Naqvi and J.~H.~Strassler,
``Giant gravitons in conformal field theory,''
JHEP \textbf{04} (2002), 034,
arXiv:hep-th/0107119.

\bibitem{CJR}
S.~Corley, A.~Jevicki and S.~Ramgoolam,
``Exact correlators of giant gravitons from dual N=4 SYM theory,''
Adv. Theor. Math. Phys. \textbf{5} (2002), 809–839,
arXiv:hep-th/0111222.

\bibitem{LLM}
H.~Lin, O.~Lunin and J.~Maldacena,
``Bubbling AdS space and 1/2 BPS geometries,''
JHEP \textbf{10} (2004), 025,
arXiv:hep-th/0409174.


\bibitem{Lynden-Bell}
D.~Lynden-Bell,
``On the negative specific heat paradox,''
Mon. Not. Roy. Astron. Soc. \textbf{181} (1977), 405.

\bibitem{HawkPage}
S.~W.~Hawking and D.~N.~Page,
``Thermodynamics of Black Holes in Anti-de Sitter Space,''
Commun. Math. Phys. \textbf{87} (1983), 577.

\bibitem{RS23} 
S. Ramgoolam and L. Sword, ``Matrix and tensor witnesses of hidden symmetry algebras'' 
JHEP 2023. 


\bibitem{CurtRein}
C.~W.~Curtis and I.~Reiner,
Representation Theory of Finite Groups and Associative Algebras,
Wiley (1962).

\bibitem{Hamermesh}
M.~Hamermesh,
Group Theory and Its Application to Physical Problems,
Addison–Wesley (1962).



\bibitem{KR}
Y.~Kimura and S.~Ramgoolam,
``Branes, anti-branes and brauer algebras in gauge-gravity duality,''
JHEP \textbf{11} (2007), 078,
arXiv:0709.2158.

\bibitem{BHR1}
T.~W.~Brown, P.~J.~Heslop and S.~Ramgoolam,
``Diagonal multi-matrix correlators and BPS operators in N=4 SYM,''
JHEP \textbf{02} (2008), 030,
arXiv:0711.0176.

\bibitem{BHR2}
T.~W.~Brown, P.~J.~Heslop and S.~Ramgoolam,
``Diagonal free field matrix correlators, global symmetries and giant gravitons,''
JHEP \textbf{04} (2009), 089,
arXiv:0806.1911.

\bibitem{BCD}
R.~Bhattacharyya, S.~Collins and R.~de~Mello~Koch,
``Exact Multi-Matrix Correlators,''
JHEP \textbf{03} (2008), 044,
arXiv:0801.2061.

\bibitem{BDS}
D.~Bekker, R.~de~Mello~Koch and M.~Stephanou,
``Exact Multi-Restricted Schur Polynomial Correlators,''
JHEP \textbf{02} (2008), 029,
arXiv:0710.5372.

\bibitem{EHS}
Y.~Kimura and S.~Ramgoolam,
``Enhanced symmetries of gauge theory and resolving the spectrum of local operators,''
Phys. Rev. D \textbf{78} (2008), 126003,
arXiv:0807.3696.


\bibitem{BHN}
V.~Balasubramanian, M.~x.~Huang, T.~S.~Levi and A.~Naqvi,
``Open strings from N=4 superYang-Mills,''
JHEP \textbf{08} (2002), 037,
arXiv:hep-th/0204196.

\bibitem{BBFH}
V.~Balasubramanian, D.~Berenstein, B.~Feng and M.~x.~Huang,
``D-branes in Yang-Mills theory and emergent gauge symmetry,''
JHEP \textbf{03} (2005), 006,
arXiv:hep-th/0411205.


\bibitem{PCA}
P.~Mattioli and S.~Ramgoolam,
Phys. Rev. D \textbf{93} (2016), 065040,
arXiv:1601.06086.


\bibitem{RobRev}
R.~de Mello Koch, M.~Kim and A.~L.~Mahu,
``A pedagogical introduction to restricted Schur polynomials with applications to heavy operators,''
Int. J. Mod. Phys. A \textbf{39} (2024) no.31, 2430003
[arXiv:2409.15751 [hep-th]].

\bibitem{BerRev}
D.~Berenstein,
``Matrix models and emergent geometry,''
PoS CORFU2017 (2018), 098
[arXiv:1806.05729 [hep-th]].



\bibitem{KempRam}
G.~Kemp and S.~Ramgoolam,
``BPS states, conserved charges and centres of symmetric group algebras,''
JHEP \textbf{01} (2020), 146,
arXiv:1911.11649.

\bibitem{QMRibb} 
J.~Ben Geloun and S.~Ramgoolam,
``Quantum mechanics of bipartite ribbon graphs: Integrality, Lattices and Kronecker coefficients,''
Alg. Comb. \textbf{6} (2023) no.2, 547-594
[arXiv:2010.04054 [hep-th]].

\bibitem{ICFDTS} 
R.~de Mello Koch, Y.~H.~He, G.~Kemp and S.~Ramgoolam,
``Integrality, duality and finiteness in combinatoric topological strings,''
JHEP \textbf{01} (2022), 071
[arXiv:2106.05598 [hep-th]].

\bibitem{RS} 
S.~Ramgoolam and E.~Sharpe,
``Combinatoric topological string theories and group theory algorithms,''
JHEP \textbf{10} (2022), 147
doi:10.1007/JHEP10(2022)147
[arXiv:2204.02266 [hep-th]].


\bibitem{ProjDetect}
J.~B.~Geloun and S.~Ramgoolam,
``The quantum detection of projectors in finite-dimensional algebras and holography,''
JHEP \textbf{05} (2023), 191
doi:10.1007/JHEP05(2023)191
[arXiv:2303.12154 [quant-ph]].

\bibitem{RowCol} 
A.~Padellaro, R.~Radhakrishnan and S.~Ramgoolam,
``Row{\textendash}column duality and combinatorial topological strings,''
J. Phys. A \textbf{57} (2024) no.6, 065202
[arXiv:2304.10217 [hep-th]].

\bibitem{PRSe} 
A.~Padellaro, S.~Ramgoolam and R.~K.~Seong,
``Row and column detection complexities of character tables,''
J. Phys. A \textbf{58} (2025) no.50, 505401
[arXiv:2503.02543 [hep-th]].


\bibitem{Hanada}
M.~Hanada and J.~Maltz,
``A proposal of the gauge theory description of the small Schwarzschild black hole in AdS$_5\times$S$^5$,''
JHEP \textbf{02} (2017), 012
[arXiv:1608.03276 [hep-th]].

\bibitem{Berenstein2018} 
D.~Berenstein,
``Negative specific heat from non-planar interactions and small black holes in AdS/CFT,''
JHEP \textbf{10} (2019), 001
[arXiv:1810.07267 [hep-th]].


\bibitem{Minwalla} 
O.~Aharony, J.~Marsano, S.~Minwalla, K.~Papadodimas and M.~Van Raamsdonk,
``The Hagedorn - deconfinement phase transition in weakly coupled large N gauge theories,''
Adv. Theor. Math. Phys. \textbf{8} (2004), 603-696
[arXiv:hep-th/0310285 [hep-th]].

\bibitem{Sundborg} 
B.~Sundborg,
``The Hagedorn transition, deconfinement and N=4 SYM theory,''
Nucl. Phys. B \textbf{573} (2000), 349-363
[arXiv:hep-th/9908001 [hep-th]].

\bibitem{RWZ}
S.~Ramgoolam, M.~C.~Wilson and A.~Zahabi,
``Quiver Asymptotics: $\mathcal{N}=1$ Free Chiral Ring,''
J. Phys. A \textbf{53} (2020), 105401,
arXiv:1811.11229.


\bibitem{BGSR}
J.~Ben~Geloun and S.~Ramgoolam,
``Counting tensor model observables and branched covers of the 2-sphere,''
Ann. Inst. H. Poincar\'e D \textbf{1} (2014), 77–138,
arXiv:1307.6490.

\bibitem{BGSR2}
J.~Ben~Geloun and S.~Ramgoolam,
``Tensor Models, Kronecker coefficients and Permutation Centralizer Algebras,''
JHEP \textbf{11} (2017), 092,
arXiv:1708.03524.


\bibitem{AsympTens}
J.~Ben Geloun and S.~Ramgoolam,
``All-orders asymptotics of tensor model observables from symmetries of restricted partitions,''
J. Phys. A \textbf{55} (2022) no.43, 435203
[arXiv:2106.01470 [hep-th]].


\bibitem{Tseyt} 
M.~Beccaria and A.~A.~Tseytlin,
``Partition function of free conformal fields in 3-plet representation,''
JHEP \textbf{05} (2017), 053
[arXiv:1703.04460 [hep-th]].

\bibitem{Kleb}
K.~Bulycheva, I.~R.~Klebanov, A.~Milekhin and G.~Tarnopolsky,
``Spectra of Operators in Large $N$ Tensor Models,''
Phys. Rev. D \textbf{97} (2018) no.2, 026016
[arXiv:1707.09347 [hep-th]]. 




\bibitem{BPR1}
Barnes, Padellaro and Ramgoolam,
``Hidden symmetries and large N factorisation for permutation invariant matrix observables,''
 JHEP \textbf{2022}, 90 (2022). 


\bibitem{BPR2}
Barnes, Padellaro and Ramgoolam, 
``Permutation symmetry in large-N matrix quantum mechanics and partition algebras,''
 Phys. Rev. D \textbf{106}, 106020 (2022)



\bibitem{PIMQM-PI} 
D.~O'Connor and S.~Ramgoolam,
JHEP \textbf{04} (2024), 080
[arXiv:2312.12397 [hep-th]].

\bibitem{PIMQM-PF} 
D.~O'Connor and S.~Ramgoolam,
``Gauged permutation invariant matrix quantum mechanics: partition functions,''
JHEP \textbf{07} (2024), 152
[arXiv:2312.12398 [hep-th]].

\bibitem{WBAPBT}
M.~Studziński, M.~Mozrzymas, P.~Kopszak and M.~Horodecki,
``Efficient multi-port-based teleportation schemes,''
Phys. Rev.A, 102, 012301 (2020),arXiv:2008.0098 4 [quant-ph].

 
\bibitem{SHOBNS} 
S.~Ramgoolam and M.~Studzi{\'n}ski,
``Simple harmonic oscillators from non-semisimple walled Brauer algebras,''
[arXiv:2509.04234 [hep-th]].



\bibitem{GAP}
The GAP Group,
\url{https://www.gap-system.org}.

\bibitem{sagemath}
The Sage Developers,
\url{https://www.sagemath.org}.




\end{thebibliography}
\end{document}